\documentclass[twoside]{ilcws10}
\usepackage[latin1]{inputenc}
\usepackage[dvips]{graphicx,epsfig,color}
\usepackage{wrapfig,rotating}
\usepackage{amssymb,amsmath,array}

\newcommand{\misspt}{$^{\mathrm{miss}}p_{\mathrm{T}}$}

\begin{document}
\title{
%%%%   Paper title goes here  %%%%%%%%%%%%%%
Identification of new physics and general WIMP search at the ILC}%% 
%***********************************************************************
% AUTHORS INFORMATION AREA
%***********************************************************************
\author{M.~Asano$^1$\thanks{e-mail: masano@tuhep.phys.tohoku.ac.jp}, 
K.~Fujii$^2$, R.~S.~Hundi$^3$, H.~Itoh$^{2,4}$, S.~Matsumoto$^5$,\\
N.~Okada$^6$, T.~Saito$^1$, T.~Suehara$^7$, Y.~Takubo$^8$, and H.~Yamamoto$^1$
% Optional short acknowledgment: remove next line if non-needed
% DO NOT MODIFY THE FOLLOWING '\vspace' ARGUMENT
\vspace{.3cm}\\
1- Tohoku University - Department of Physics \\
6-3 Aoba, Aramaki, Aoba-ku, Sendai, Miyagi, 980-8578 - Japan
\vspace{.1cm}\\
2- High Energy Accelerator Research Organization - Institute of Particle and Nuclear Studies \\
1-1 Oho, Tsukuba, Ibaraki, 305-0801 - Japan
\vspace{.1cm}\\
3- University of Hawaii - Department of Physics \& Astronomy \\
2505 Correa Rd.~Honolulu, HI, 96822 - United States
\vspace{.1cm}\\
4- The University of Tokyo - Institute of Cosmic Ray Research \\
5-1-5 Kashiwa-no-ha, Kashiwa, Chiba, 277-8582 - Japan
\vspace{.1cm}\\
5- University of Toyama - Department of Physics, Graduate School of Science and Engineering \\
3190 Gofuku, Toyama, Toyama, 930-8555 - Japan
\vspace{.1cm}\\
6- The University of  Alabama - Department of Physics \& Astronomy \\
Tuscaloosa, AL, 35487-0324 - United States
\vspace{.1cm}\\
7- The University of Tokyo - International Center for Elementary Particle Physics \\
7-3-1 Hongo, Bunkyo-ku, Tokyo, 113-0033 - Japan
\vspace{.1cm}\\
8- Tohoku University - Center for the Advancement of Higher Education \\
41 Kawauchi, Aoba-ku, Sendai, Miyagi, 980-8576 - Japan
}
%%***********************************************************************
% END OF AUTHORS INFORMATION AREA
%***********************************************************************

\maketitle

\begin{abstract}
We investigate the possibility of the identification of TeV physics models 
including WIMP dark matter at the International Linear Collider. Many TeV 
physics models contain a WIMP dark matter ($\chi^0$) and charged new particle 
($\chi^{\pm}$) which interacts with the WIMP dark matter via the vertex 
$\chi^{\pm} \chi^0 W^{\mp}$. Through Monte Carlo simulations, we study 
the process, $e^+e^- \rightarrow \chi^+ \chi^- \rightarrow \chi^0 \chi^0 W^+ W^- $,
because the signal contains the fruitful information of the model. We show that, 
in particular, the distribution of the $\chi^{\pm}$ production angle is the 
powerful probe in the TeV physics model search. 
\end{abstract}

\section{Introduction}
About 23 $\%$ energy density of the present Universe is made up of dark 
matter and it plays a key role in the large structure formation. 
%Dark matter also play a key role in the large structure formation.
However, we don't know what it is. Dark matter candidates don't exist 
in the standard model (SM). 

%For reason given later, 
Weakly Interacting Massive Particle (WIMP) is one of the most plausible candidate 
for dark matter. It is neutral, massive, and sufficiently stable particle and 
such particle naturally explains the observed dark matter abundance,
$\Omega_{DM}h^2 \sim 0.11$ \cite{Komatsu:2010fb}.
Many models of physics beyond the SM including WIMP dark matter have been proposed.
%It also seems plausible that WIMP dark matter is in the TeV physics model which 
%explains the smallness of electroweak scale 
%because the new particles would interact with the SM particles weakly and 
%the masses are $\mathcal{O}(100)$ GeV. 
%If the lightest new particle is neutral and significantly stable 
%(e.g. due to the R-parity in supersymmetric(SUSY) models), it is a WIMP candidate.
% 

%We are interested in how to measure the WIMP dark matter nature.
The TeV physics could be produced 
at the Large Hadron Collider (LHC) and the International Linear Collider (ILC). 
The LHC is now operating, but the precision measurement of events including 
dark matter in final state is difficult at the experiments. On the other hand, 
it would be possible at the ILC.
The measurements is important for not only dark matter physics but also 
identification of the TeV new physics model.

In this study, we investigate the possibility of general WIMP search 
and new physics model identification studying
 $e^+e^- \rightarrow \chi^+ \chi^- \rightarrow \chi^0 \chi^0 W^+ W^- $ 
process at the ILC, where $\chi^0$ and $\chi^{\pm}$ denote the WIMP and the 
charged new particle, respectively. 
Using the process, we have already studied about the possibility of precision 
measurements of the Little Higgs model \cite{Asakawa:2009qb}. We have shown 
that we can extract fruitful information from the signal of the process: 
The masses of $\chi^{\pm}$ and $\chi^0 $ can be determined from the edges of 
the energy distribution of the reconstructed $W$ bosons. It is also possible 
to confirm that the spin of $\chi^{\pm}$ and the vertex structure using the 
angular distribution of the $\chi^{\pm}$ pair procudion and the polarization 
of $W^{\pm}$. The gauge charges of the $\chi^{\pm}$ boson could 
be measured using a polarized electron beam.

It is worth noting that such process exists in many TeV physics models 
which explain the smallness of electroweak scale , because WIMP candidates would 
interact with the SM particles weakly. Thus, the process plays a 
key role of the general WIMP search.

%%%%%%%%%%%%%%%%%%%%%%%%%%%%%%%%%%%%%%%%%%%%%%%%%%%%%%%%%%%%%%%%%%%%%%%%%%%%%%%%%
\begin{center}
\begin{table*}[t]
  \begin{tabular}{|l||c|c|}
  \hline
  Models & Particles & Spins \\
  \hline
  Inert Higgs like model & ($\chi^{\pm}_S$, $\chi^0_S$) &  (0, 0)   \\
  \hline
  SUSY like model & ($\chi^{\pm}_F$, $\chi^0_F$) &  (1/2, 1/2)   \\
  \hline
  Little Higgs like model & ($\chi^{\pm}_V$, $\chi^0_V$) &  (1, 1)   \\
  \hline
  (1, 0) model & ($\chi^{\pm}_V$, $\chi^0_S$) &  (1, 0)   \\
  \hline
  (0, 1) model & ($\chi^{\pm}_S$, $\chi^0_V$) &  (0, 1)   \\
  \hline
  \end{tabular}
  \caption{\small Classification of spin configuration of new particles.}
  \label{tab:spin_combi}
\end{table*}
\end{center}
%%%%%%%%%%%%%%%%%%%%%%%%%%%%%%%%%%%%%%%%%%%%%%%%%%%%%%%%%%%%%%%%%%%%%%%%%%%%%%%%%

%%%%%%%%%%%%%%%%%%%%%%%%%%%%%%%%%%%%%%%%%%%%%%%%%%%%%%%%%%%%%%%%%%%%%%%%%%%%%%%%%%%%%%%%%%
\section{Benchmark models and representative points}
%%%%%%%%%%%%%%%%%%%%%%%%%%%%%%%%%%%%%%%%%%%%%%%%%%%%%%%%%%%%%%%%%%%%%%%%%%%%%%%%%%%%%%%%%%

We concentrate on the WIMP dark matter($\chi^0$) and the charged new particle 
($\chi^{\pm}$) which interacts with dark matter. The interaction vertex, 
$\chi^{\pm} \chi^0 W^{\mp} $, is also exist in many TeV physics models which 
contain WIMP scenario. All spin combinations which has the vertex is 
written in Table \ref{tab:spin_combi}, keeping charge, gauge, discrete 
symmetry and Lorentz invariance. 

In this study, we investigate 
the Inert Higgs doublet like model (IH) \cite{Barbieri:2006dq}, 
supersymmetric like model (SUSY) \cite{BookDrees}, 
and Littlest Higgs like model (LHT) \cite{ArkaniHamed:2001nc}-\cite{Hubisz:2004ft}
as benchmark models of $\chi^{\pm}$ spin 0, 1/2, and 1, respectively
%%
%\footnote{ 
%For the general $Z\chi^{+}\chi^{-}$ vertex is discussed in last section.
%}
%%
. The crucial difference from the (1,0) and (0,1) models in 
Table \ref{tab:spin_combi} only appears in the what relates with the 
$\chi^{\pm}\chi^{0}W^{\mp}$ vertex 
(e.g. the shape of the energy distribution of W bosons).

At representative points, we take the same new particle masses and 
production cross section in each models 
in order to investigate the 
%possibility of the model identification 
separation possibility of the TeV new physics model
using the same masses and number of events. 
Actually, when we discuss the 
features of new particles which observed at future colliders, we compare the 
signal with the sample models in the same way.
The parameters at the representative points are summarized in Table \ref{tab:sample points}. 

Since the mass difference of new particles is determined in littlest Higgs model
explicitly, the mass differences at representative points are adjusted to 
coincidence with Littlest Higgs model. We take the coefficients of vertices to 
consistent with a parameter in each models which realizes the mass difference. 
However, in order to adjust the amplitude of cross section to each other, the over all of
neutral couplings (e.g. $\chi^{+} \chi^{-} Z$) are normalized.

In this paper, we show the study at $\sqrt{s}=1$TeV with an integrated 
luminosity of 500 fb$^{-1}$. The study at $\sqrt{s}=500$GeV is written in 
\cite{Suehara:2010rd}.

%%%%%%%%%%%%%%%%%%%%%%%%%%%%%%%%%%%%%%%%%%%%%%%%%%%%%%%%%%%%%%%%%%%%%%%%%%%%%%%%%
\begin{center}
\begin{table*}[t]
  \begin{tabular}{|c||c|c|c|c|}
  \hline
  & $m_{\chi^{\pm}}$ [GeV] & $m_{\chi^{0}}$ [GeV] 
  & Cross section 
  & Br($\chi^{\pm} \to \chi^{0} W^{\pm} $) \\
  \hline
  $\sqrt{s} = 500$ GeV & 231.57 & 44.03 
  & 40, 200 [fb$^{-1}$] & $\sim 100 \%$ \\
  \hline
  $\sqrt{s} = 1$   TeV & 368 & 81.9 
  & 40, 200 [fb$^{-1}$] &    $\sim 100 \%$ \\
  \hline
  \end{tabular}
  \caption{\small Representative points in this study}
  \label{tab:sample points}
\end{table*}
\end{center}
%%%%%%%%%%%%%%%%%%%%%%%%%%%%%%%%%%%%%%%%%%%%%%%%%%%%%%%%%%%%%%%%%%%%%%%%%%%%%%%%%
%%%%%%%%%%%%%%%%%%%%%%%%%%%%%%%%%%%%%%%%%%%%%%%%%%%%%%%%%%%%%%%%%%%%%%%%%%%%%%%%%%%%%%%%%%
\section{Simulation tools}
%%%%%%%%%%%%%%%%%%%%%%%%%%%%%%%%%%%%%%%%%%%%%%%%%%%%%%%%%%%%%%%%%%%%%%%%%%%%%%%%%%%%%%%%%%

In our study, both signal and background events have been generated 
by Physsim \cite{physsim}. The initial-state radiation and beamstrahlung have been included 
in the event generations. 
%The beam energy spread was set to XXX\%. 
We have ignored the finite crossing angle between the electron and positron beams. 
In the event generations, helicity amplitudes were calculated using the HELAS library \cite{helas}, 
which allows us to deal with the effect of gauge boson polarizations properly. 
Phase space integration and the generation of parton 4-momenta have been performed by 
BASES/SPRING \cite{bases}. Parton showering and hadronization have been carried out by using 
PYTHIA6.4 \cite{pythia}, where final-state tau leptons are decayed by TAUOLA \cite{tauola} 
in order to handle their polarizations correctly.

%%% description for 1TeV %%%%
The generated Monte Carlo events have been passed to a detector simulator called JSFQuickSimulator, 
which implements the GLD geometry and other detector-performance related parameters \cite{glddod}. 
In the detector simulator, hits by charged particles at the vertex detector and track parameters 
at the central tracker are smeared according to their position resolutions, taking into account 
correlations due to off-diagonal elements in the error matrix. Since calorimeter signals are 
simulated in individual segments, a realistic simulation of cluster overlapping is possible. 
Track-cluster matching is performed for the hit clusters in the calorimeter in order to achieve 
the best energy flow measurements. The resultant detector performance in our simulation study is 
summarized in Table \ref{tb:GLD}.

\begin{table}
 \center{
  \begin{tabular}{lcr}
   \hline
   Detector & Performance & Coverage \\
   \hline
   Vertex detector &
   $\delta_{b} \leq 5 \oplus 10/ p \beta \sin^{3/2}\theta$ ($\mu$m) &
   $|\cos\theta| \leq 0.93$
   \\
   Central drift chamber &
   $\delta p_{t}/p_{t}^{2} \leq 5 \times 10^{-5}$ (GeV/c)$^{-1}$ &
   $|\cos\theta| \leq 0.98$
   \\
   EM calorimeter &
   $\sigma_{E}/E = 17\% / \sqrt{E} \oplus 1\%$ &
   $|\cos\theta| \leq 0.99$
   \\
   Hadron calorimeter &
   $\sigma_{E}/E = 45\% / \sqrt{E} \oplus 2\%$ &
   $|\cos\theta| \leq 0.99$
   \\
   \hline
  \end{tabular}
 }
 \caption{\small Detector parameters used in our simulation study.}
 \label{tb:GLD}
\end{table}

%%%%%%%%%%%%%%%%%%%%%%%%%%%%%%%%%%%%%%%%%%%%%%%%%%%%%%%%%%%%%%%%%%%%%%%%%%%%%%%%%%%%%%%%%%
\section{Signal Selection}
%%%%%%%%%%%%%%%%%%%%%%%%%%%%%%%%%%%%%%%%%%%%%%%%%%%%%%%%%%%%%%%%%%%%%%%%%%%%%%%%%%%%%%%%%%

We use events in which both $W$ bosons decay hadronically, 
 $e^+e^- \to \chi^+ \chi^- \to \chi^0 \chi^0 W^+ W^- \to \chi^0 \chi^0 qqqq$, 
as signal events. Thus, the main background processes are $W^+W^-$,$\nu \bar{\nu}W^+W^-$, etc.

In order to identify the two $W$ bosons from $\chi^{\pm}$ decays, two jet-pairs 
have been selected so as to minimize a $\chi^2$ function,
\begin{equation}
\chi^2
= 
(^{\mathrm{rec}}\mathrm{M}_{W1} -~^{\mathrm{tr}}\mathrm{M}_{W})^{2}/\sigma_{\mathrm{M}_{W}}^{2} 
+ 
(^{\mathrm{rec}}\mathrm{M}_{W2} -~^{\mathrm{tr}}\mathrm{M}_{W})^{2}/\sigma_{\mathrm{M}_{W}}^{2},
\end{equation}
where $^{\mathrm{rec}}\mathrm{M}_{W1(2)}$ is the invariant mass of the first 
(second) 2-jet system paired as a $W$ candidate, $^{\mathrm{tr}}\mathrm{M}_{W}$ 
is the true $W$ mass (80.4 GeV), and $\sigma_{\mathrm{M}_{W}}$ is the resolution 
for the $W$ mass (4 GeV). We required $\chi^2 < 26$ to obtain well-reconstructed events. 

Since $\chi^{0}$ escape from detection resulting in a missing momentum, we have 
thus selected events with the missing transverse momentum \misspt~above 84 GeV. 
We have also selected events with a energy of $W$ below 500 GeV.
The number of remaining background events is much smaller than that of the signal 
after imposing all the cuts. 

%\begin{table}
%\center{
%\begin{tabular}{|l|r|r|r|}
%\hline
%Process  & Cross sec. [fb] & \# of events & \# of events after all cuts \\
%\hline
%\ccqqqqnn   & 91.9  & 45,968  & (IH) 32,944 \\
%            &       &         & (SUSY) 32,101 \\
%            &       &         & (LHT) 31,214 \\ \hline 
%\wwqqqq     & 1,773 & 886,500 & 39 \\ 
%\enwzenqqqq & 25.5  & 12,750  & 3,714 \\
%\nnwwnnqqqq & 6.45  & 3,225   & 1,473 \\
%\eewweeqqqq & 465   & 232,500 & 3 \\
%\hline
%\end{tabular}
%}
%\caption{Cut summary for $\sqrt{s} = 1$ TeV. The cross-section and the number of events of \ccqqqqnn~are set to XXX fb for all the physics models in this table.}
% \label{tb:cut_summary1tev}
%\end{table}

%%%%%%%%%%%%%%%%%%%%%%%%%%%%%%%%%%%%%%%%%%%%%%%%%%%%%%%%%%%%%%%%%%%%%%%%%%%%%%%%%%%%%%%%%%
\section{Mass Determination}
%%%%%%%%%%%%%%%%%%%%%%%%%%%%%%%%%%%%%%%%%%%%%%%%%%%%%%%%%%%%%%%%%%%%%%%%%%%%%%%%%%%%%%%%%%

The Masses of new particles, $\chi^{0}$ and $\chi^{\pm}$, can be determined from 
the edges of the $W$ energy distribution. After subtracting the backgrounds, the 
distribution has been fitted with a line shape determined by a high statistics 
signal sample. The fitted masses of $\chi^{0}$ and $\chi^{\pm}$ are summarized 
in Table \ref{tb:reso_1tev}. 
The masses of $\chi^{\pm}$ and $\chi^{0}$ will be determined with $\mathcal{O}(1)$\%
accuracy.

\begin{table}
\center{
\begin{tabular}{|l|r|r|r|}
\hline
 & Physics model & $\sigma = 40$ fb & $\sigma = 200$ fb \\
\hline
$M_{\chi^{\pm}}$ (GeV) & IH   & 364.3 $\pm$ 6.0 & 366.4 $\pm$ 1.4 \\
                       & SUSY & 370.6 $\pm$ 5.6 & 368.0 $\pm$ 1.3 \\
                       & LHT  & 367.7 $\pm$ 4.0 & 367.4 $\pm$ 0.9 \\ \hline 
$M_{\chi^{0}}$ (GeV)   & IH   & 79.7 $\pm$ 4.8 & 78.5 $\pm$ 1.2 \\
                       & SUSY & 76.6 $\pm$ 6.5 & 77.8 $\pm$ 1.3 \\
                       & LHT  & 78.0 $\pm$ 3.9 & 78.7 $\pm$ 0.9 \\ 
\hline
\end{tabular}
}
\caption{Measurement accuracy of masses of $\chi^{\pm}$ and $\chi^{0}$.}
 \label{tb:reso_1tev}
\end{table}

%%%%%%%%%%%%%%%%%%%%%%%%%%%%%%%%%%%%%%%%%%%%%%%%%%%%%%%%%%%%%%%%%%%%%%%%%%%%%%%%%%%%%%%%%%
\section{Angular Distribution}
%%%%%%%%%%%%%%%%%%%%%%%%%%%%%%%%%%%%%%%%%%%%%%%%%%%%%%%%%%%%%%%%%%%%%%%%%%%%%%%%%%%%%%%%%%

\begin{figure}[thb]
% 	\begin{minipage}[t]{.24\textwidth}
% 	\begin{center}
% 		\includegraphics[width=0.95\columnwidth]{ntrack.eps}
% 		(i)\\[1cm]
% 	\end{center}
% 	\end{minipage}
	\begin{minipage}[t]{.32\textwidth}
	\begin{center}
		\includegraphics[width=1\columnwidth, angle=270]{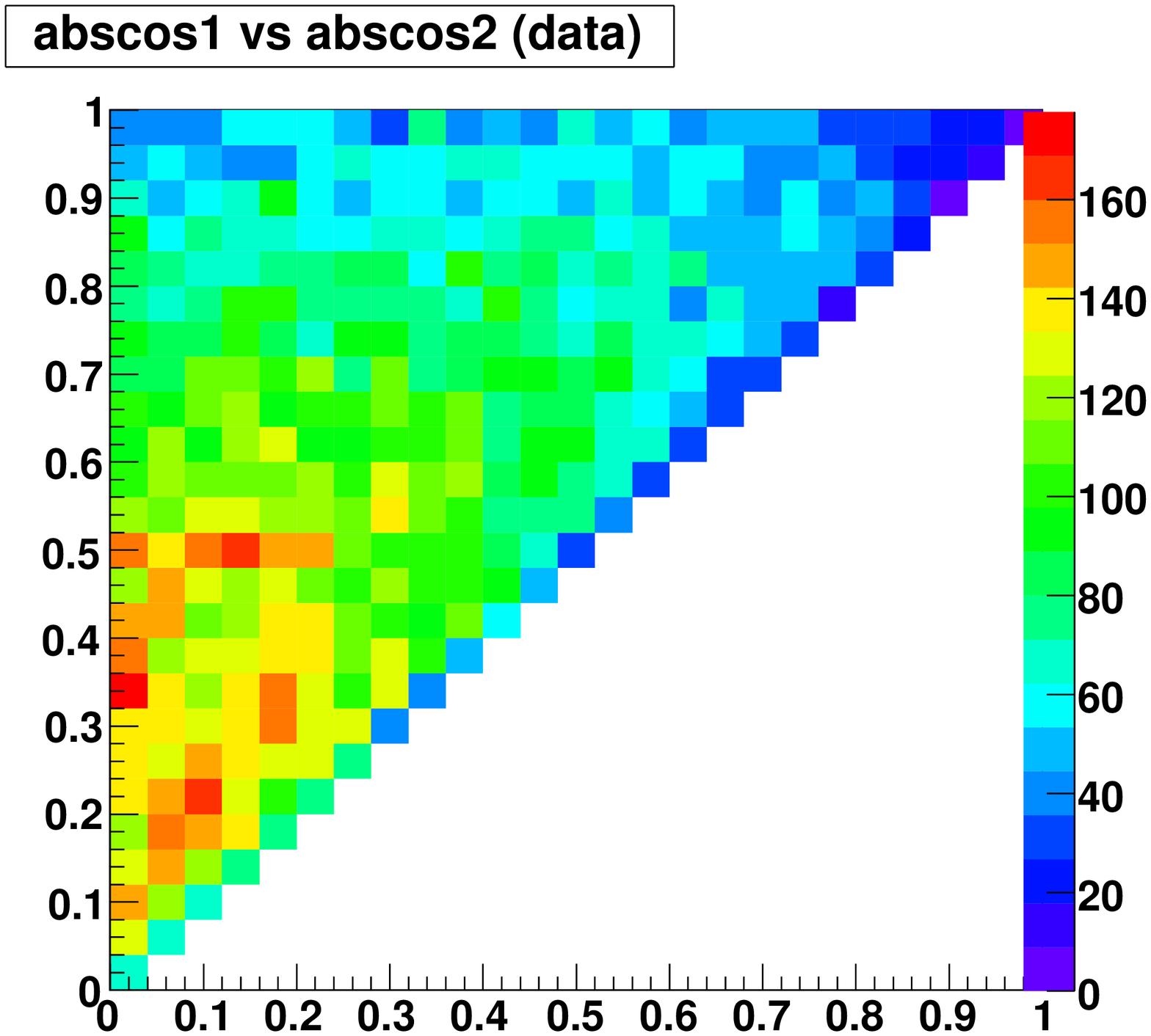}
		(IH)\\[1cm]
	\end{center}
	\end{minipage}
	\begin{minipage}[t]{.32\textwidth}
	\begin{center}
		\includegraphics[width=1\columnwidth, angle=270]{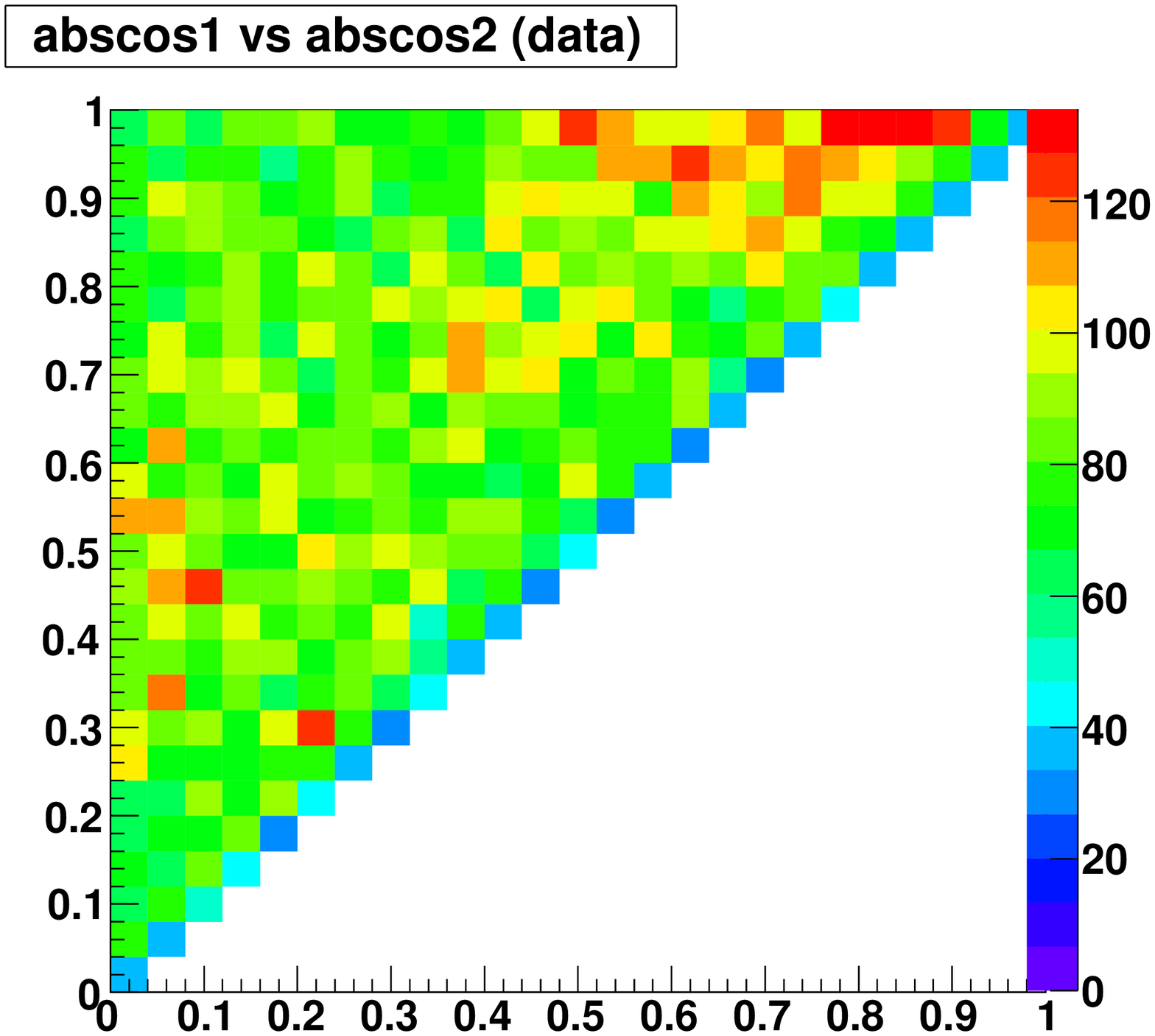}
		(SUSY)\\[1cm]
	\end{center}
	\end{minipage}
	\begin{minipage}[t]{.32\textwidth}
	\begin{center}
		\includegraphics[width=1\columnwidth, angle=270]{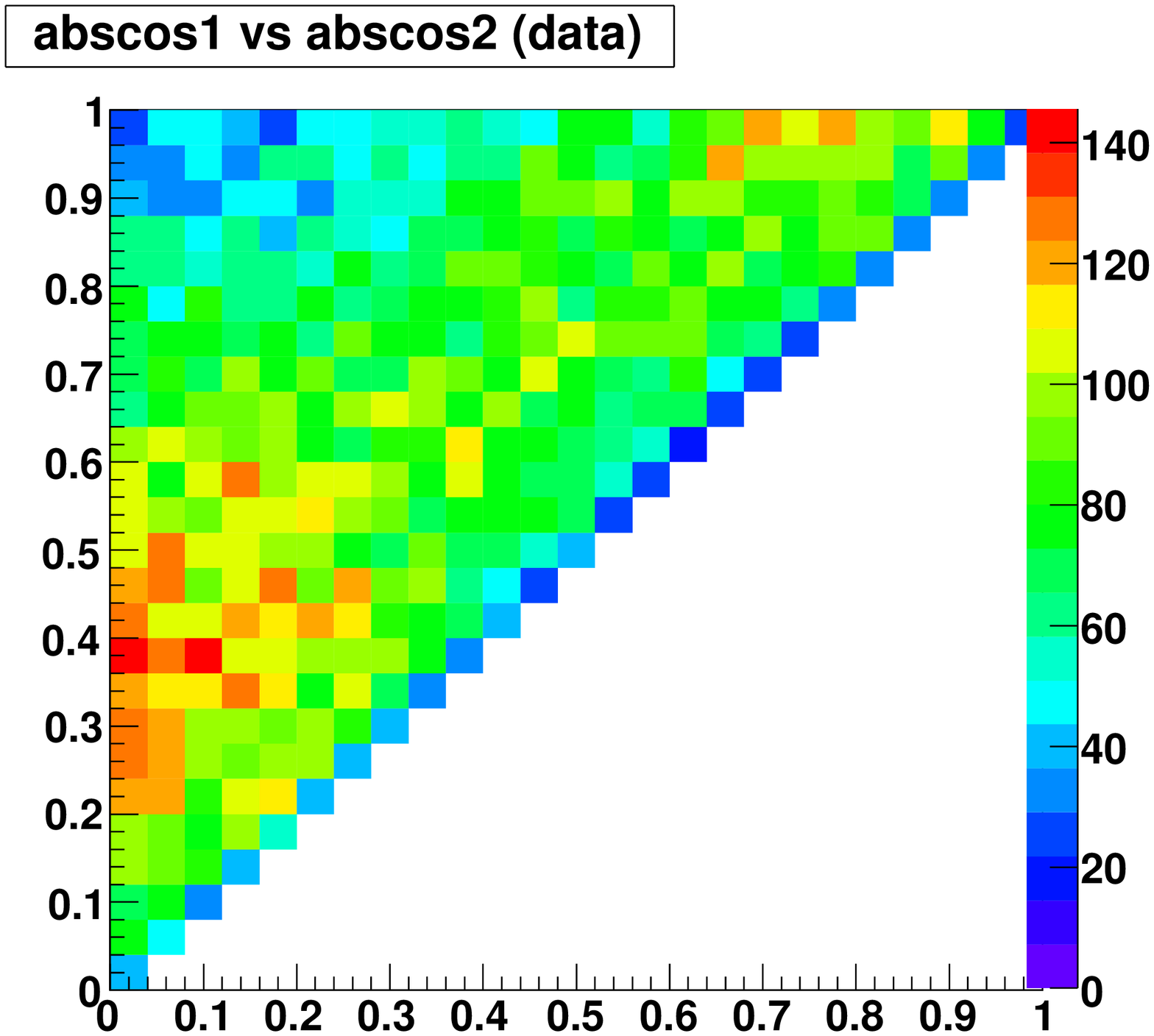}
		(LH)\\[1cm]
	\end{center}
	\end{minipage}
	\caption{The 2-dimensional histogram for the two solutions of the two solutions of the $\chi^{\pm}$ production angle at each models.}
	\label{fig:angle}
\end{figure}

The production angle of $\chi^{\pm}$ can be calculated with 2-fold ambiguity from 
the momenta of $W$ bosons, assuming back-to-back production of $\chi^{+}$ and $\chi^{-}$.
Figure \ref{fig:angle}
%\ref{}
shows 2-dimensional histogram. 
To estimate the possibility of the new physics model identification,
we compare the distribution with the "template" which is the distribution of 
high statistics signal samples using $\chi^{2}$ analysis: 
%--------------------------------------------------------->>>
\begin{eqnarray}
\chi^2
&=& \frac{N_{\rm Data} - N_{\rm Temp}}{\sigma_{\rm Sig + BG}},
    \label{fig:xi}
\end{eqnarray}
%---------------------------------------------------------<<<
where $N_{\rm Data}$ and $N_{\rm Temp}$ are the number of the data and 
the number of the template at each bin, respectively. The $\sigma_{\rm Sig + BG}$ 
is the error of the signal and backgrounds. The $\chi^{2}$ values are summarized in 
Table \ref{tb:chi2_1tev}. It shows that the identification of these new physics models 
is possible by comparing of production angles of $\chi^{\pm}$.

\begin{table}
\center{
\begin{tabular}{|l|r|r|r|r|}
\hline
$\sigma$  & Model & IH template & SUSY template & LHT template \\
\hline
40 fb   &  IH & 1.2 & 4.7 & 3.6 \\
       & SUSY & 5.2 & 1.2 & 3.2 \\
       & LHT  & 3.7 & 2.6 & 1.2 \\ \hline 
200 fb &  IH  & 1.0 & 17.3 & 11.0 \\
       & SUSY & 20.0 & 1.0 & 9.2 \\
       & LHT  & 13.7 & 7.8 & 1.0 \\ 
\hline
\end{tabular}
}
\caption{Summary of $\chi^{2}/$(the number of bins) values from 2-dimensional histograms 
of the $\chi^{\pm}$ production angle.}
 \label{tb:chi2_1tev}
\end{table}

%%%%%%%%%%%%%%%%%%%%%%%%%%%%%%%%%%%%%%%%%%%%%%%%%%%%%%%%%%%%%%%%%%%%%%%%%%%%%%%%%%%%%%%%%%
\section{Summary}
%%%%%%%%%%%%%%%%%%%%%%%%%%%%%%%%%%%%%%%%%%%%%%%%%%%%%%%%%%%%%%%%%%%%%%%%%%%%%%%%%%%%%%%%%%

We study the possibility of general WIMP search and new physics model identification 
using $e^+e^- \rightarrow \chi^+ \chi^- \rightarrow \chi^0 \chi^0 W^+ W^- $ 
process at the $\sqrt{s} = 1$ TeV ILC with integrated luminosity of 500 fb$^{-1}$. 
We have shown that the masses of new particles can be determined very accurately at the ILC 
in a model independent way. Furthermore, using $\chi^2$ analysis, the identification of new 
physics models is possible by comparing of the $\chi^{\pm}$ production angles. 
Finally, we have also studied the threshold scan to separate the new physics model.
It shows that the SUSY like case will be separate from other models.
%As a result, it is separate the SUSY case from other models. 

\section*{Acknowledgments}
The authors would like to thank all members of the ILC physics subgroup \cite{ILC_phys_subgroup}
for useful discussions. This work is supported in part by the Grant-in-Aid for the Global
COE Program Weaving Science Web beyond Particle-matter Hierarchy from the Ministry of Education, 
Culture, Sports, Science and Technology of Japan and 
the Creative Scientific Research Grant (No. 18GS0202) of the Japan Society for Promotion of Science 
and the JSPS Core University Program.

%This study is supported in part by the Creative Scientific Research Grant No.~18GS0202 of the
%Japan Society for Promotion of Science.

% ****************************************************************************
% BIBLIOGRAPHY AREA
% ****************************************************************************

\begin{footnotesize}
% IF YOU DO NOT USE BIBTEX, USE THE FOLLOWING SAMPLE SCHEME FOR THE REFERENCES
% ----------------------------------------------------------------------------

% ----------------------------------------------------------------------------

\end{footnotesize}

% ****************************************************************************
% END OF BIBLIOGRAPHY AREA
% ****************************************************************************


\begin{thebibliography}{99}
% Please replace the numbers for   contribId   and   sessionId
% in the following URL. You can get this information by going to 
% http://indico.cern.ch/confAuthorIndex.py?confId=2628
% and search for your contribution and click on the title
% Be aware: '&amp;' must be replaced by simple '&' as in example below
%------- replace following references ;-)
%%% Intro %%%
\bibitem{Komatsu:2010fb}
  E.~Komatsu {\it et al.},
  %``Seven-Year Wilkinson Microwave Anisotropy Probe (WMAP) Observations:
  %Cosmological Interpretation,''
  arXiv:1001.4538 [astro-ph.CO].
  %%CITATION = ARXIV:1001.4538;%%

\bibitem{Asakawa:2009qb}
  E.~Asakawa {\it et al.},
  %``Precision Measurements of Little Higgs Parameters at the International
  %Linear Collider,''
  Phys.\ Rev.\  D {\bf 79} (2009) 075013
  [arXiv:0901.1081 [hep-ph]].
  %%CITATION = PHRVA,D79,075013;%%

%%% MODELs %%%

%--- IH ---%
\bibitem{Barbieri:2006dq}
  R.~Barbieri, L.~J.~Hall and V.~S.~Rychkov,
  %``Improved naturalness with a heavy Higgs: An alternative road to LHC
  %physics,''
  Phys.\ Rev.\  D {\bf 74} (2006) 015007
  [arXiv:hep-ph/0603188].
  %%CITATION = PHRVA,D74,015007;%%

%--- SUSY ---%
\bibitem{BookDrees}
See, for example,
M.~Drees, R.~M.~Godbole and P.~Roy,
{\it Theory and phenomenology of sparticles},
(World Scientific, 2004).

%--- LHT ---%
\bibitem{ArkaniHamed:2001nc}
  N.~Arkani-Hamed, A.~G.~Cohen and H.~Georgi,
  %``Electroweak symmetry breaking from dimensional deconstruction,''
  Phys.\ Lett.\  B {\bf 513} (2001) 232
  [arXiv:hep-ph/0105239].
  %%CITATION = PHLTA,B513,232;%%

\bibitem{ArkaniHamed:2002qx}
  N.~Arkani-Hamed, A.~G.~Cohen, E.~Katz, A.~E.~Nelson, T.~Gregoire and J.~G.~Wacker,
  %``The Minimal Moose for a Little Higgs,''
  JHEP {\bf 0208} (2002) 021
  [arXiv:hep-ph/0206020].
  %%CITATION = JHEPA,0208,021;%%

\bibitem{ArkaniHamed:2002qy}
  N.~Arkani-Hamed, A.~G.~Cohen, E.~Katz and A.~E.~Nelson,
  %``The littlest Higgs,''
  JHEP {\bf 0207} (2002) 034
  [arXiv:hep-ph/0206021].
  %%CITATION = JHEPA,0207,034;%%

\bibitem{Cheng:2003ju}
  H.~C.~Cheng and I.~Low,
  %``TeV symmetry and the little hierarchy problem,''
  JHEP {\bf 0309} (2003) 051
  [arXiv:hep-ph/0308199].
  %%CITATION = JHEPA,0309,051;%%

\bibitem{Cheng:2004yc}
  H.~C.~Cheng and I.~Low,
  %``Little hierarchy, little Higgses, and a little symmetry,''
  JHEP {\bf 0408} (2004) 061
  [arXiv:hep-ph/0405243].
  %%CITATION = JHEPA,0408,061;%%

\bibitem{Low:2004xc}
  I.~Low,
  %``T parity and the littlest Higgs,''
  JHEP {\bf 0410} (2004) 067
  [arXiv:hep-ph/0409025].
  %%CITATION = JHEPA,0410,067;%%

\bibitem{Hubisz:2004ft}
  J.~Hubisz and P.~Meade,
  %``Phenomenology of the littlest Higgs with T-parity,''
  Phys.\ Rev.\  D {\bf 71} (2005) 035016
  [arXiv:hep-ph/0411264](For the correct
paramter region consistent with the WMAP observation, see the figure in the
revised vergion, hep-ph/0411264v3).
  %%CITATION = PHRVA,D71,035016;%%

%%% Proceeding %%%
%\cite{Suehara:2010rd}
\bibitem{Suehara:2010rd}
  T.~Suehara {\it et al.},
  %``Simulation Study of W Boson + Dark Matter Signatures for Identification of
  %New Physics,''
  arXiv:1007.0829 [hep-ex].
  %%CITATION = ARXIV:1007.0829;%%

%%% Simulation tool %%%
\bibitem{physsim} http://acfahep.kek.jp/subg/sim/softs.html.
\bibitem{helas} H. Murayama, I. Watanabe, K. Hagiwara, KEK-91-11, (1992) 184.
\bibitem{bases} T. Ishikawa, T. Kaneko, K. Kato, S. Kawabata,\emph{Comp, Phys. Comm.} {\bf 41} (1986) 127.
\bibitem{pythia} T. Sj$\dot{\mathrm{o}}$strand, \emph{Comp, Phys. Comm.} {\bf 82} (1994) 74.
\bibitem{tauola} http://wasm.home.cern.ch/wasm/goodies.html.
\bibitem{glddod} GLD Detector Outline Document, arXiv:physics/0607154.


\bibitem{ILC_phys_subgroup}  
 http://www-jlc.kek.jp/subg/physics/ilcphys/.\\
\end{thebibliography}
\end{document}